\documentstyle[twoside,fleqn,espcrc2,psfig]{article}

\newcommand{\NP}[1]{ Nucl.\ Phys.\ {\bf #1}}

\newcommand{\PL}[1]{ Phys.\ Lett.\ {\bf #1}}

\newcommand{\PR}[1]{Phys.\ Rev.\ {\bf #1}}
\newcommand{\PRL}[1]{ Phys.\ Rev.\ Lett.\ {\bf #1}}

\newcommand{\AmS}{{\protect\the\textfont2
  A\kern-.1667em\lower.5ex\hbox{M}\kern-.125emS}}

\title{Perturbative QCD relations inspired
by hypothetical tau leptons        
\thanks{Research partially supported by the Spanish 
CICYT under contract AEN97-1693
and the U.S. Department of Energy DE-AC03-76SF00515.}
}

\author{
J.R.Pel\'aez\address{Departamento de F\'{i}sica Te\'orica.
Universidad Complutense de Madrid.\\
28040 Madrid. SPAIN},
S.J.Brodsky \address{Stanford Linear Accelerator Center. Stanford University.\\
Stanford, California 94309. U.S.A.},
C.Merino \address{Departamento de F\'{\i}sica de Part\'{\i}culas. 
Universidade de Santiago de Compostela.\\ 15706 Santiago de Compostela. SPAIN}
and N. Toumbas  \address{Stanford Linear Accelerator Center. Stanford University.\\
Stanford, California 94309. U.S.A.}
}

\begin{document}

\begin{abstract} We review our recent works on
 tests of perturbative QCD, inspired by the relation 
between the hadronic decay of the
$\tau$ lepton and the  $e^+ e^-$
annihilation into hadrons.
First, we present a set of
commensurate scale relations that probe the
self-consistency of leading-twist QCD predictions
for any observable which defines an effective charge. These tests
are independent of the renormalization scheme and scale,
and are applicable over wide data ranges.
As an example we apply this approach to $R_{e^+ e^-}$.
Second, using a differential form of these conmensurate scale
relations, we present a method to measure the QCD Gell-Mann--Low
$\Psi$ function.
\end{abstract}

\maketitle

\pagestyle{empty}
\arraycolsep=0.5pt
\section{Introduction}

The  hadronic width of the $\tau$ lepton,
$R_{\tau} = \Gamma(\tau^-\rightarrow{\nu_{\tau}
+{\rm hadrons}})/\Gamma(\tau^-\rightarrow{\nu_{\tau}e^-\bar{\nu_e}})$
plays an important role in
the determination of the QCD coupling \cite{Braaten}. For
this analysis it is essential to 
minimize the sensitivity to low energy data
\cite{Davier} using different integral moments. 
In addition, by integrating the data on
spectral functions up $M$ we can simulate
the physics of hypothetical $\tau$ leptons\cite{Davier} with smaller
$M$ than the real one.
Since  their hadronic widths are related to the $e^+e^-$ annihilation
cross section into hadrons $R_{e^+e^-}$ through
\begin{eqnarray}
&&R_{\tau}(M) =
\frac{2}{(\sum_f{q_f^2})}
\times\label{defRt}
\\\nonumber
&&\int^{M^2}_0\frac{\,d\,s}{M^2}
\left(1 -
\frac{s}{M^2}\right)^2\left(1 +
\frac{2s}{M^2}\right)R_{e^+e^-}(\sqrt{s}).
\end{eqnarray}
they provide tests of perturbative QCD (PQCD).

Our aim in this paper is to report
on several different applications motivated by the above relation,
where we proposed self-consistency tests of PQCD,
applicable to any observable which
defines an effective charge \cite{nosotros},
as well as a method to measure the Gell-Mann--Low function
of some of these observables \cite{Carlos}.

An effective charge defined from an observable  
contain its entire radiative
contribution \cite{effcha}. For instance,
assuming $f$ massless flavors, we have
\begin{eqnarray}
R_{e^+e^-}(\sqrt{s})&\equiv&R_0
\left[1+\frac{\alpha_R(\sqrt{s})}{\pi}\right],\nonumber\\
R_{\tau}(M)&\equiv& R^0_{\tau}(M)
\left[1+\frac{\alpha_\tau(M)}{\pi}\right],
\label{ReePQCD}
\end{eqnarray}
where $R_0=3\sum_f
q_f^2$ The effective charges $\alpha_R$ and $\alpha_\tau$ can
be written as a series
in $\alpha_s/\pi$ in any given renormalization scheme.
In addition we  demand that effective charges should also be
linear in $\alpha_s$, such as $\Delta R=R_{e^+e^-}-R_0$,
and it also must
track the bare charge in its coupling to the various quark flavors.
(Thus $R_{e^+e^-}^2 - R^2_0 = (R_{e^+e^-}+R_0) \Delta R$ 
is unacceptable since it is not
linear in quark flavor \footnote {We thank Prof. M.A. Braun for 
discussions on this point}).
Their relevance is due to the fact that they
satisfy the renormalization group equation with the same
coefficients $\beta_0$ and $\beta_1$ as 
$\alpha_s$.

As a consequence of the Mean Value Theorem applied to
eq.(\ref{defRt}), $\alpha_R$ and $\alpha_\tau$ 
are simply related by a scale shift
$\alpha_\tau(M) = \alpha_R(\sqrt{s^*})$. However
we can use NLO leading twist QCD to {\em predict}
\begin{equation}
\lambda_\tau=
\frac{\sqrt{s^*}}{M} = \exp\left[ -{19\over 24} - {169\over 128}
{\alpha_R(M)\over \pi} + \cdots\right],
\label{CSStau}
\end{equation}
This result was first obtained in \cite{CSR} by using the NNLO,
and it is a particular example of a ``commensurate scale relation''
\cite{CSR} which relate observables at two different scales.
We will now see that it is due to the fact that both effective
charges evolve with universal
$\beta_0$ and $\beta_1$ coefficients \cite{nosotros}.

\section{Tests of PQCD for a general observable}

The previous relations can be generalized 
to an arbitrary observable
$O(\sqrt{s})$, with an associated effective
charge $\alpha_O$, by defining new effective charges
\begin{equation}
\alpha_f(M) \equiv \frac{\int^{M^2}_0 \frac{d\,s}{M^2}\,f\left(
\frac{s}{M^2}\right)\alpha_O(\sqrt{s})}{\int^{M^2}_0
\frac{d\,s}{M^2}f\left(
\frac{s}{M^2}\right)},
\label{alphaRf}
\end{equation}
where we can choose $f(x)$ to be any smooth, integrable function of
$x=s/M^2$. Once again
\begin{eqnarray*}
\alpha_f(M)=\alpha_O(\sqrt{s^*_f}),\hspace{2cm}    0\leq s^*_f\leq M^2.
\end{eqnarray*}
Let us remark
that this relation {\em involves only data for 
$O(\sqrt{s})$} and is therefore a self-consistency test
of the leading twist QCD approximation. The relation between
the commensurate scales is obtained if 
we consider the running of $\alpha_O$ up to third
order
\begin{eqnarray*}
\frac{\alpha_O(\sqrt{s})}{\pi}= \frac{\alpha_O(M)}{\pi} -
\frac{\beta_0}{4} \ln{\left(\frac{s}{M^2}\right)}
\left(\frac{\alpha_O(M)}{\pi}\right)^2 +\\\nonumber
\frac{1}{16}\left[
\beta_0^2 \ln^2\left(\frac{s}{M^2}\right)-
\beta_1 \ln\left(\frac{s}{M^2}\right)\right]
\left(\frac{\alpha_O(M)}{\pi}\right)^3,
\end{eqnarray*}
and we substitute for $\alpha_O$
in eq. (\ref{alphaRf}), to find
\begin{eqnarray*}
\frac{\alpha_f(M)}{\pi} = \frac{\alpha_O(M)}{\pi} -
\frac{\beta_0}{4}\left(\frac{I_1}{I_0}\right)
\left(\frac{\alpha_0(M)}{\pi}\right)^2
+ \nonumber\\
\frac{1}{16}\left[\beta_0^2 \left(\frac{I_2}{I_0}\right)-
\beta_1 \left(\frac{I_1}{I_0}\right)\right]
\left(\frac{\alpha_O(M)}{\pi}\right)^3,
\end{eqnarray*}
where  $I_l = \int^{1}_0f(x)(\ln{x})^ld\,x$. Thus,
$\lambda_f \equiv \frac{\sqrt{s^*_f}}{M}$ is

\begin{equation}
\lambda_f=\exp{\left[\frac{I_1}{2I_0} +
\frac{\beta_0}{8}\left(\left(\frac{I_1}{I_0}\right)^2 - \frac{I_2}{I_0}
\right)\frac{\alpha_O(M)}{\pi}\right]}.
\label{sstar}
\end{equation}
Since $\lambda_f$ is
constant to leading order, $\alpha_f$ satisfies 
the same renormalization group
equation as $\alpha_O$ with the same coefficients $\beta_0$ and
$\beta_1$; i.e., $\alpha_f$ is an effective charge.

\subsection{Example: $R_{e^+e^-}$ data.}

For illustrative purposes, we will
set $O=R_{e^+e^-}$, since the data is
known to present problems in some energy ranges, but
a good agreement with QCD in others. We will see that our
test is able to detect these problematic regions.
In order to suppress the low energy effects,
where non-perturbative effects are important,
we shall set $f(x)=x^k$,
with $k$ some positive number. Thus
\begin{equation}
\alpha_k(M) = \alpha_R(\lambda_k\,M)\quad \hbox{with}\quad
\lambda_k=e^{\frac{-1}{2(1+k)}},
\label{atk}
\end{equation}
When comparing with $R_{e^+e^-}$ data,
we take into account finite mass effects using \cite{PQW}:
\begin{eqnarray}
R_{e^+e^-}(\sqrt{s})&=&3\sum_1^f q_i^2 T(v_i) \left[ 1+g(v_i)
\frac{\alpha_R(\sqrt{s})}{\pi}\right]
\nonumber \\
&\equiv& R_0(\sqrt{s})+R_{Sch}(\sqrt{s})\frac{\alpha_R(\sqrt{s})}{\pi}
\label{Rsch} \\
g(v)&=&\frac{4\pi}{3}\left[\frac{\pi}{2v}-\frac{3+v}{4}\left(
\frac{\pi}{2}-\frac{3}{4\pi}\right)\right]\label{g(v)}
\end{eqnarray}
where $v_i$ is the velocity of the initial
quarks in their CM frame. The $T(v_i)=v_i(3-v_i^2)/2$ factor is the parton model
mass dependence and $g(v)$ is a QCD modification of
the Schwinger correction.
The quark masses have been taken as effective parameters which
provide a good fit to the smeared data.
These corrections spoil eq.({\ref{atk}}), but
they are only important near the quark thresholds. Since
we recover eq.(\ref{ReePQCD}) at higher energies, our
study is restricted to this regime.
\begin{figure}[htb]
\psfig{file=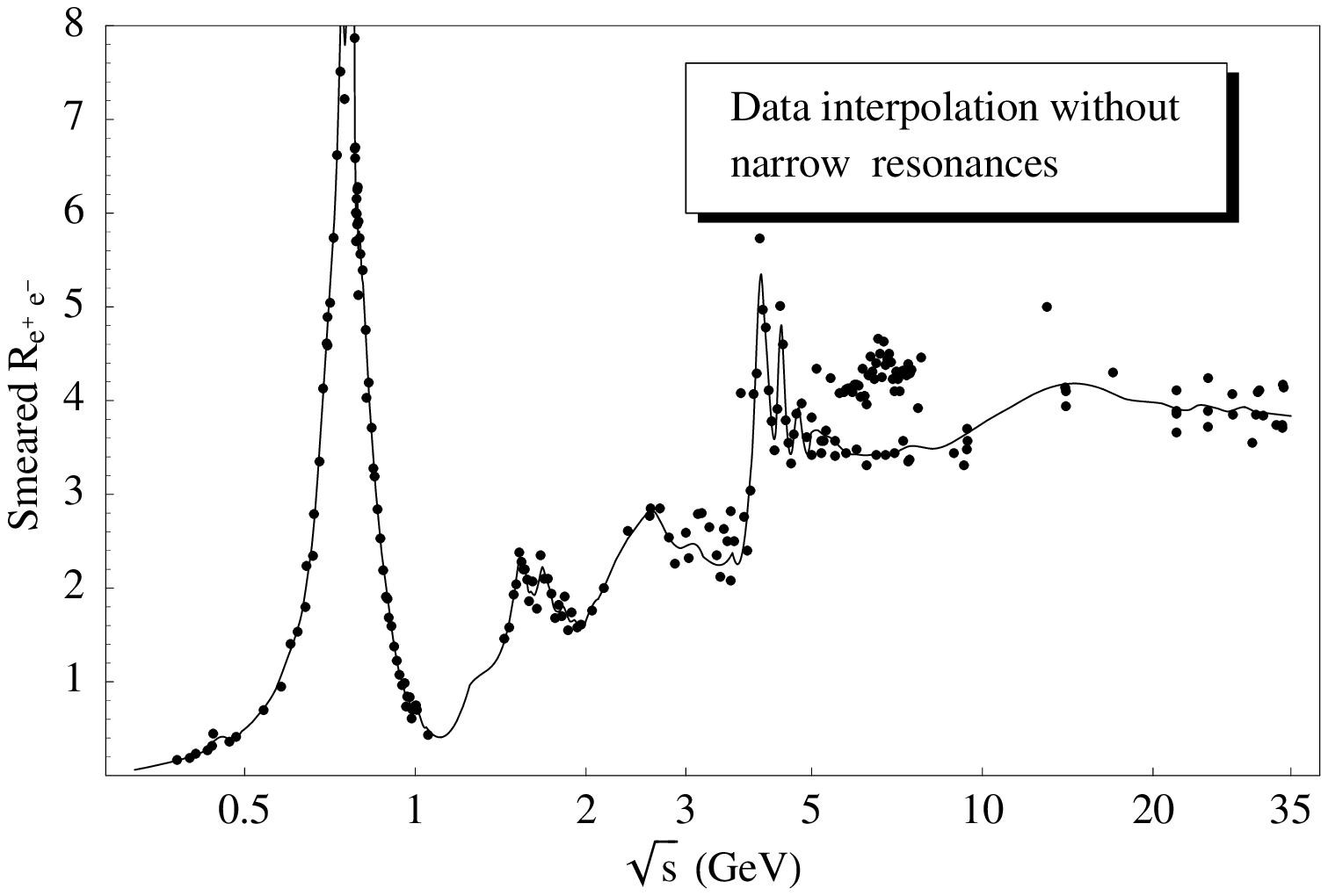,width=7.5cm}

\psfig{file=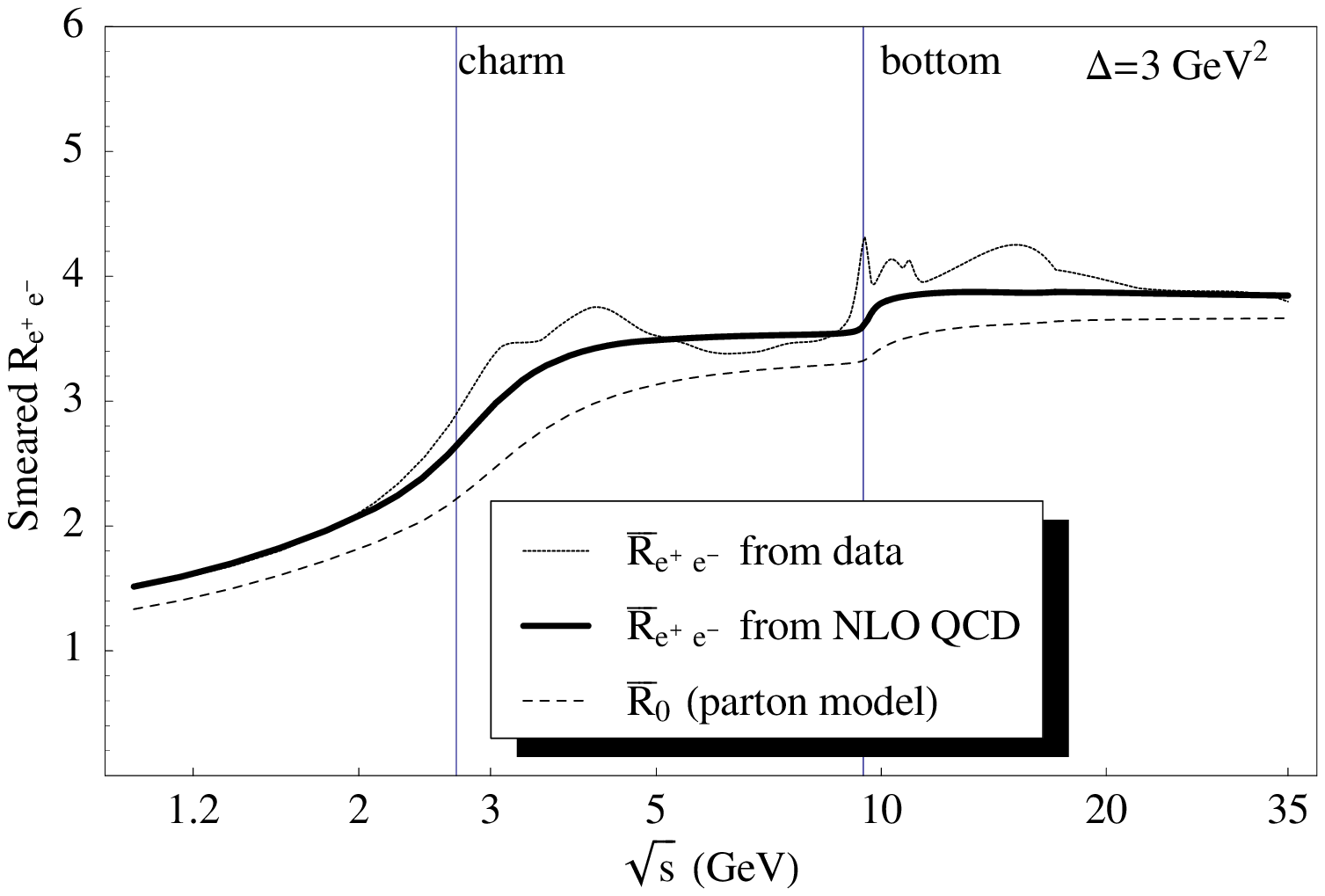,width=7.5cm}
{\footnotesize {\bf Figure 1.a)}
Interpolation of the central values of $R_{e^+e^-}$ data (see \cite{nosotros}
for references). Note the discrepancy in the central values of experiments
 between 5 and 10 GeV.
{\bf b)} Smeared $R_{e^+e^-}$.}
\end{figure}
Since the data is on annihilation into hadrons, not quarks,
we cannot use directly our formulae, although
 we can define smeared quantities \cite{PQW} as follows:
\begin{equation}
\bar{R}(\sqrt{s})=
\frac{\Delta}{\pi}\int^\infty_0 \frac{R(\sqrt{s'})}
{(s-s')^2-\Delta^2}\,d\,s'
\label{smear}
\end{equation}
The smearing of $R_{e^+e^-}$ over a range of energy,
$\Delta E$, is indeed a smearing over time $\Delta t = 1/\Delta
E $, where an analysis in terms of quarks and gluons
is appropriate. We have used the standard value $\Delta=3\,\hbox{GeV}^2$
\cite{PQW,MaSt}. The smearing effect can be seen 
by comparing the data interpolation in Fig.1.a,
(see \cite{nosotros} for references) with Fig. 1.b.
Note that {\em any fit using the QCD functional 
dependence will  always satisfy}
eq.(\ref{atk}). That is why we have parameterized
the narrow resonances with their Breit-Wigner form, and we have interpolated
the remaining data (see \cite{nosotros} for details).

Finally, using eqs.(\ref{Rsch}) and (\ref{smear}), we define
smeared charges:
\begin{equation}
\bar{\alpha}_R(\sqrt{s})=
\frac{\bar{R}_{e^+e^-}(\sqrt{s})-\bar{R}_0(\sqrt{s})}{\bar{R}_{Sch}(\sqrt{s})},
\end{equation}
and similarly for $\bar{\alpha}_k$.
As we have just commented, the smeared charges are expected to satisfy
eq.(\ref{sstar}) at energies where the threshold corrections are negligible.

In Fig.2.a we show the comparison between $\bar{\alpha}_R(\sqrt{s^*})$ and
$\bar{\alpha}_k (\sqrt{s^*}/\lambda_k)$.
The rather poor agreement for
$\alpha_0$  is due to the fact that the low energy region
is not suppressed enough. In contrast
there is a fairly good agreement for $\alpha_1$ in several regions,
which disappears when the scales are not shifted as in eq.(\ref{atk}).
As we have commented there are two regions of particular interest:
First, from 5 to 10 GeV, since the data from two different
experiments are incompatible (see Fig.1.a and
ref.\cite{discrepancy}). Even though we have kept the most recent data
in our calculations, their central values
are systematically lower than the QCD predictions.
Our test correctly points out this incompatibility.
\begin{figure}[hbt]

\psfig{file=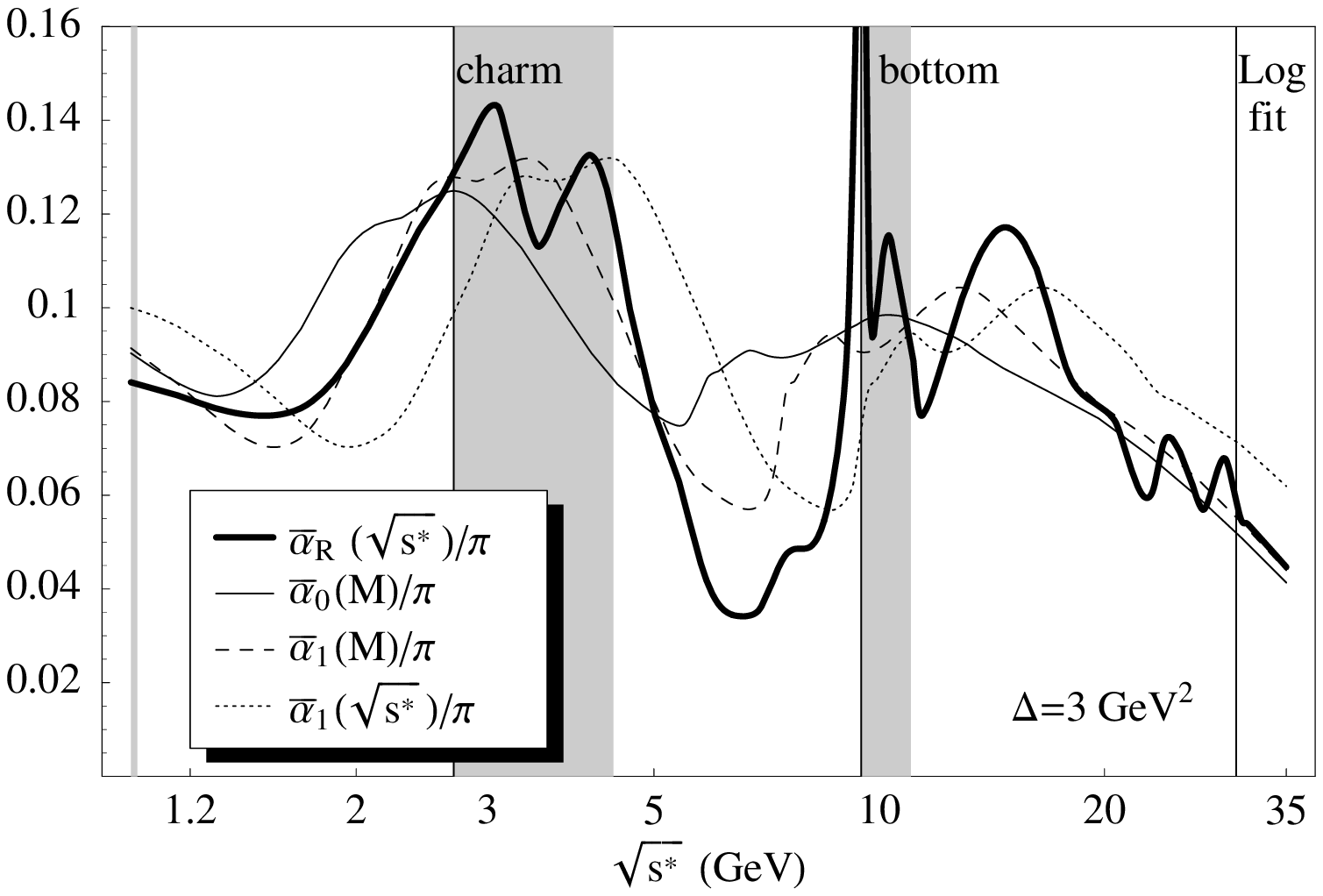,width=7.5cm}

\psfig{file=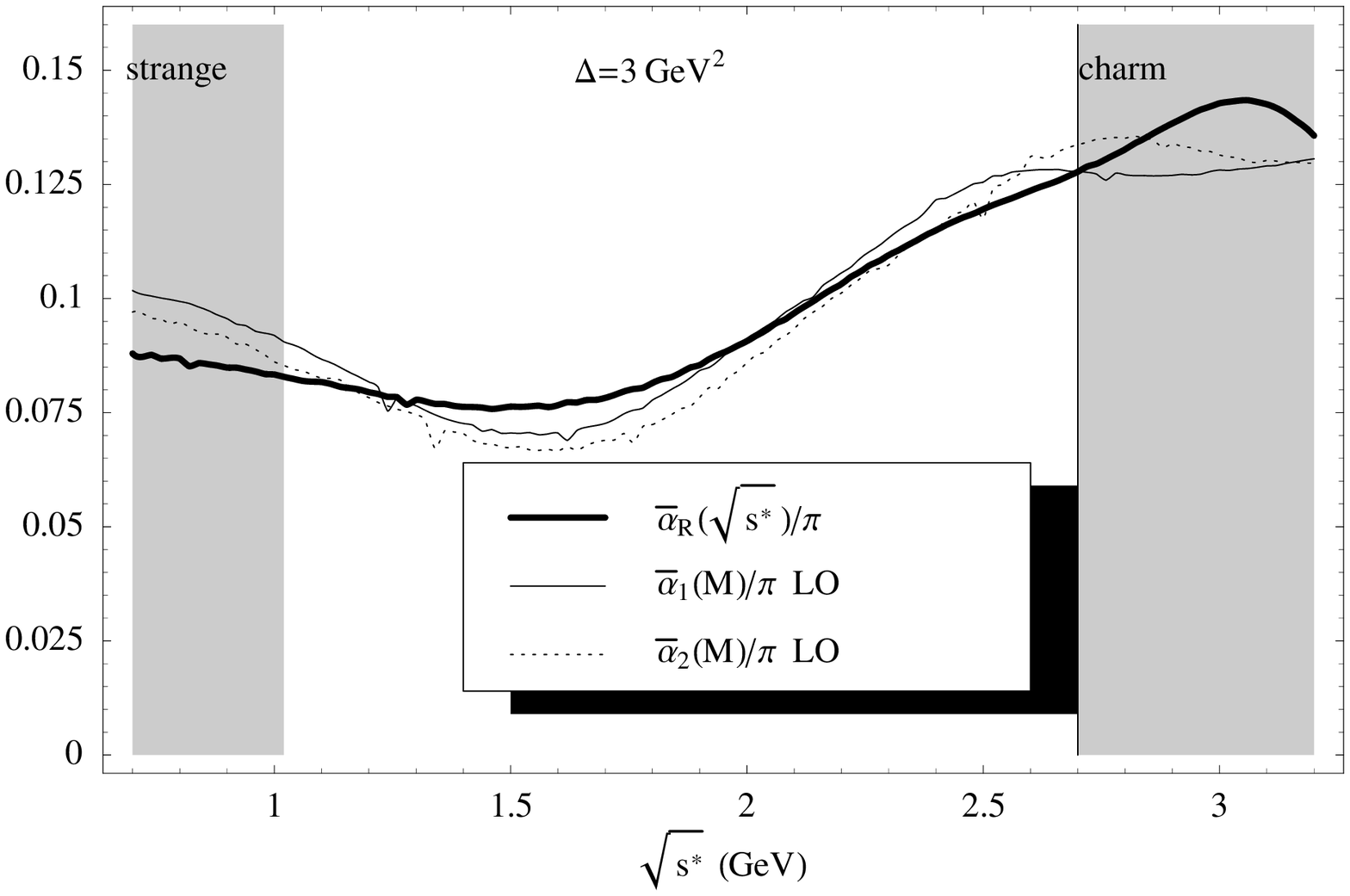,width=7.5cm}

{\footnotesize {\bf Figure 2.a)}
Comparison between $\bar\alpha_R(\sqrt{s^*})$ and different
$\bar{\alpha}_k$ moments at $M=\sqrt{s^*}/\lambda_k$.
The dotted line shows how the agreement is spoilt
if we do not shift $\sqrt{s^*}$ to $M$.
{\bf b)}
Comparison between $\bar\alpha_R(\sqrt{s^*})$ and different
$\bar{\alpha}_k$ moments at $M=\sqrt{s^*}/\lambda_k$
in the low energy region.
}
\end{figure}

Second, we show in Fig. 2.b. the energy region of the
real $\tau$. taking into account 
that we are just using LO QCD and central data values,
the agreement is fairly good, which 
supports the applicability of PQCD at energies
near the physical $\tau$ lepton.
Nevertheless, at $\sqrt{s}\sim 1.5$ GeV,
it seems that the there is an small discrepancy of the
order of 6-7\%, which also supports previous claims that
$R_{e^+e^-}$ data is lower than 
 expected from $R_\tau$ data
\cite{Groote}.

Note, however, that we have reached our conclusions 
using {\em only data on $R_{e^+e^-}$}. 

\section{Measuring the QCD Gell-Mann--Low $\Psi$ function.}
Given an observable $O(\sqrt{s})$, let us generalize the 
definition of new effective charges to include
arbitrary scales in the integral
  \begin{eqnarray}
O_{k}(\Lambda) 
\equiv C\int_{\Lambda_1^2}^{\Lambda_2^2}
\frac{d\,s}{\Lambda^2}\left(\frac{s}{\Lambda^2}\right)^{k}
O(\Lambda).
\end{eqnarray}
The generalization of eq.(\ref{alphaRf})
to obtain an effective charge $\alpha_{k}$ is straightforward.
Again we find a commensurate scale relation
$\alpha_{k}(\Lambda)= \alpha_{O}(\Lambda_{k})$, with
$\Lambda_k$ {\em predicted} by QCD \cite{Carlos}. But
if we differentiate
\begin{eqnarray}
&&\frac{d\,O_{k}(\Lambda)}{d\,\log\Lambda}=\\\nonumber
&&2C\left[\lambda_2^{2k+2}O(\Lambda_2)-
\lambda_1^{2k+2}O(\Lambda_1)\right]-(2k+2)
O_{k}(\Lambda)
\end{eqnarray}
which is basically $\beta_{QCD}$. Note that we have
defined $\lambda_i=\Lambda_i/\Lambda$.
Indeed what we have obtained is the Gell-Mann--Low
QCD function $\Psi$ (see ref.\cite{Carlos}),
which finally can be written 
in terms of {\em the observable  at three energies}
\begin{eqnarray}
\beta_{QCD}(\Lambda)&&\simeq
\Psi(\Lambda)=\\ \nonumber
&&\frac{k+1}{O^0}\left[\frac{\lambda_2^{2k+2}
O(\Lambda_2)-O(\Lambda)}
{\lambda_2^{2k+2}-1}
-O(\Lambda_k)\right].
\end{eqnarray}
The advantages of this 
 method are that
it eliminates the errors from finite difference approximations,
and that it is competitive with a standard fit but it has 
very different systematics.

\section{Summary}

The relation between $R_{e^+e^-}$ and $R_\tau$, as well as
the ideas of commensurate scale relations, have motivated
a number of new tests of perturbative QCD, as well the means
for measuring the QCD $\Psi$ Gell-Mann--Low function.
The advantage of these methods is that, although they
are inspired in a relation between two
observables, at the end they only relate one observable with itself.

These tests, which are renormalization scheme and scale independent,
 are applicable over wide
energy ranges to any observable which defines an effective charge.
As an example, we have tested the self-consistency of
existing $R_{e^+e^-}$ data according to PQCD, finding
 a good agreement in the real $\tau$ region but  
incompatibilities around the
1.5 GeV region and in the range of 5 to 10 GeV, 
supporting previous claims
in the literature.

The method to measure the Gell-Mann--Low
function eliminates the errors present in finite
difference techniques, and it is competitive with standard fits
although with different systematics.

\end{document}